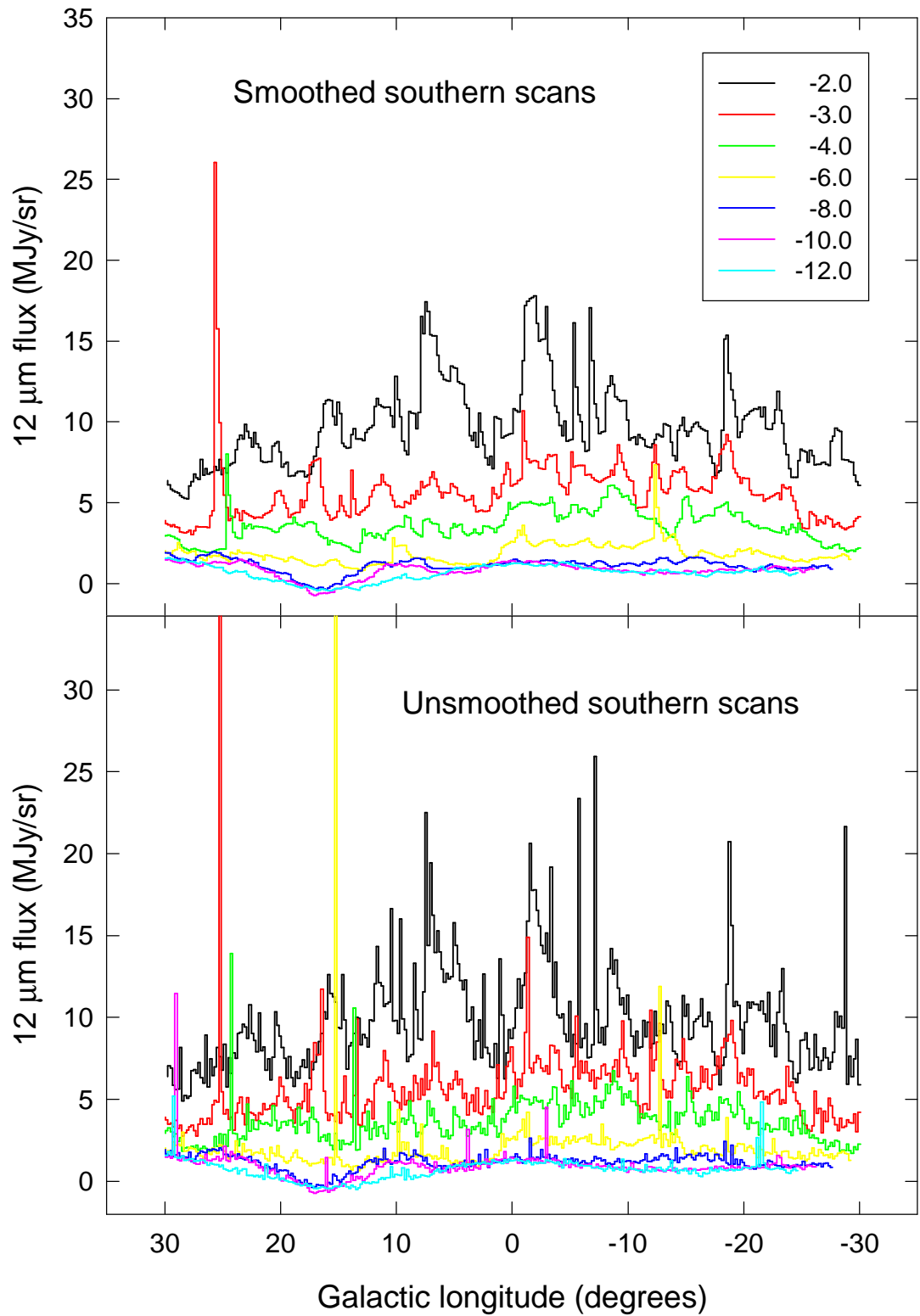

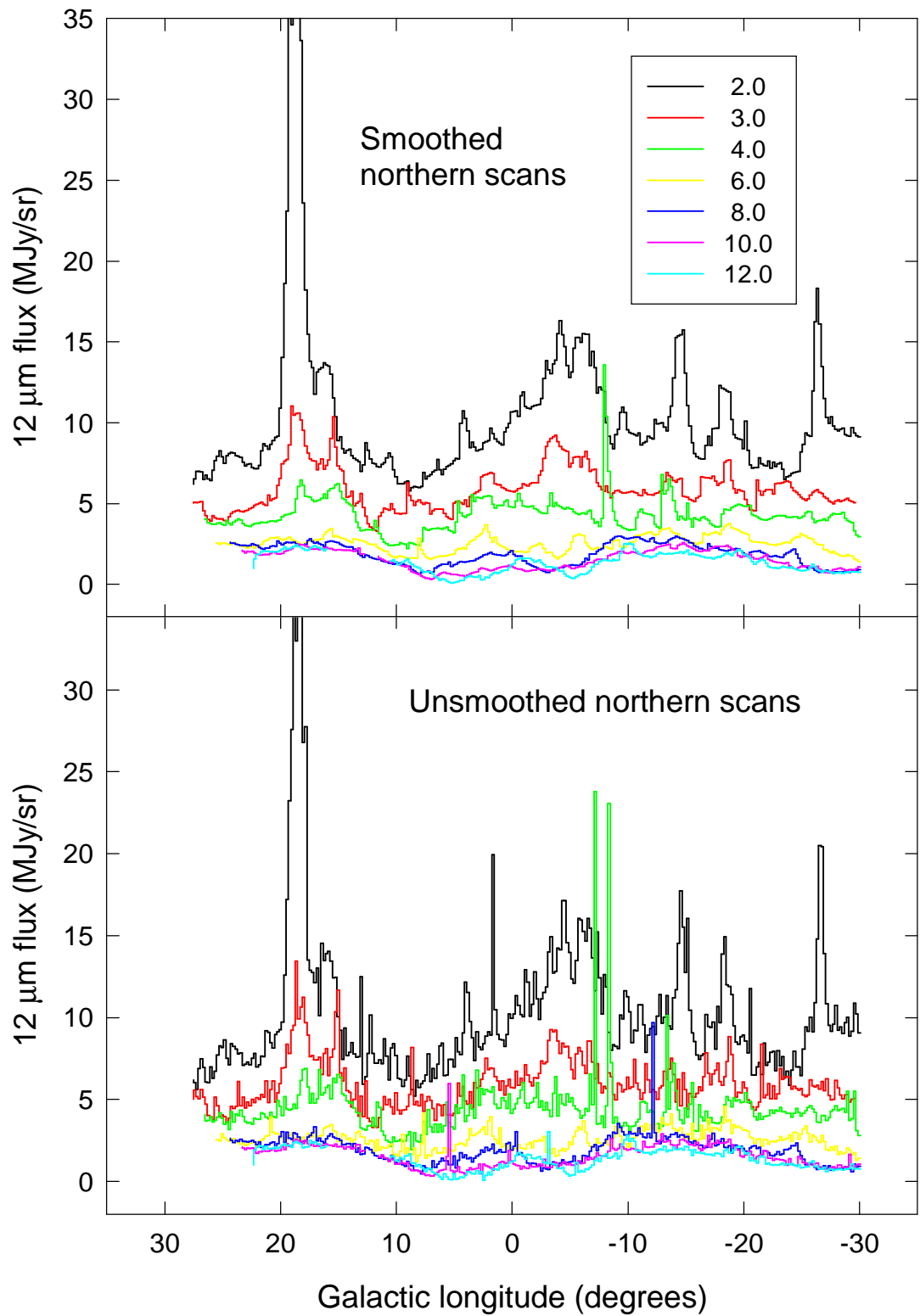

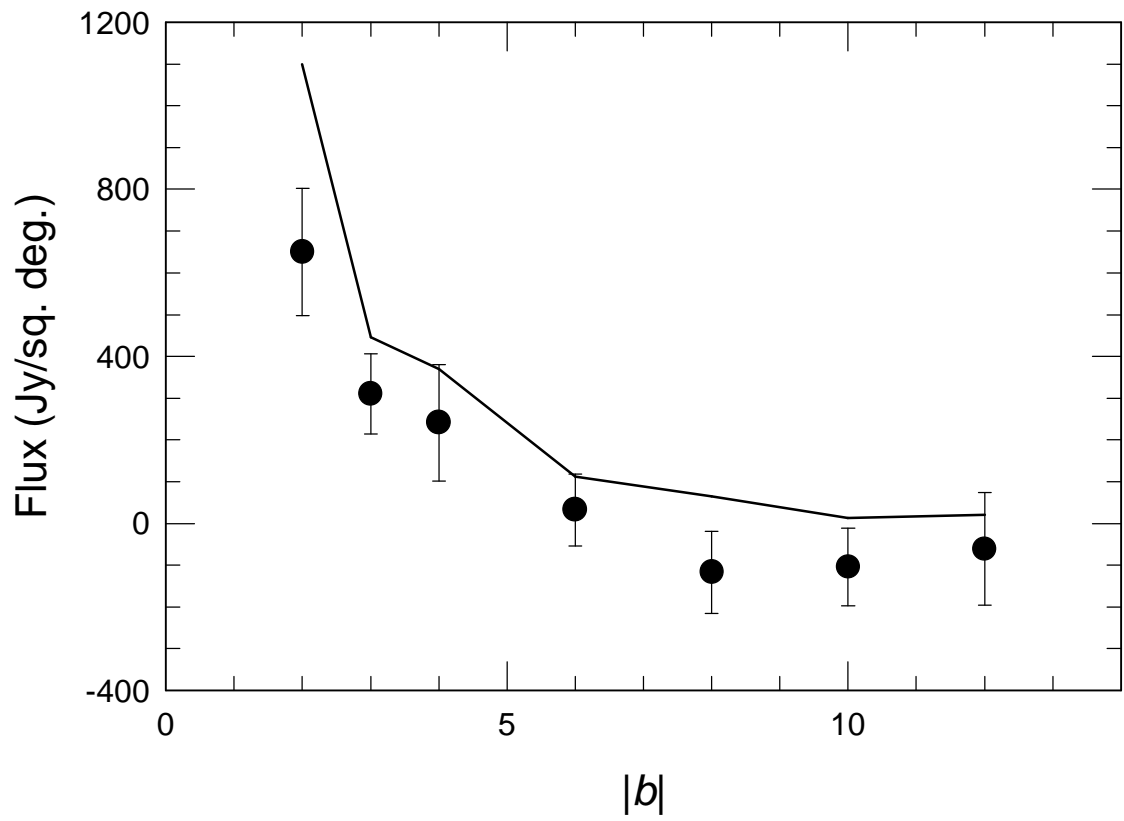

Comparison of observations with model

# An Attempt to Detect the
# Galactic Bulge at 12 μm with IRAS

*Jay A. Frogel*

Department of Astronomy, The Ohio State University

174 West 18th Avenue, Columbus, Ohio  43210

e-mail:  frogel@galileo.mps.ohio-state.edu

## ABSTRACT

Surface brightness maps at 12 µm, derived from observations with the Infrared Astronomical Satellite (IRAS), are used to estimate the integrated flux at this wavelength from the Galactic bulge as a function of galactic latitude along the minor axis. A simple model was used to remove Galactic disk emission (*e.g.* unresolved stars and dust) from the IRAS measurements. The resulting estimates are compared with predictions for the 12 µm bulge surface brightness based on observations of complete samples of optically identified M giants in several minor axis bulge fields. No evidence is found for any significant component of 12µm emission in the bulge other than that expected from the optically identified M star sample plus normal, lower luminosity stars. Known large amplitude variables and point sources from the IRAS catalogue contribute only a small fraction to the total 12 µm flux.

# 1. INTRODUCTION AND BACKGROUND

The bolometric luminosity of early-type galaxies and of the bulges of most spiral galaxies is dominated by emission from old, cool giants (Aaronson 1977; Frogel *et al.* 1978; Frogel 1988). The rapid luminosity evolution of these giants (Tinsley 1972), which include asymptotic giant branch (AGB) stars, combined with their scarcity in the solar neighborhood, has made it difficult to quantitatively assess their contribution to stellar synthesis models. Yet, a determination of the so-called "passive" luminosity evolution of a galaxy, i.e., that which results solely from the evolution of its stellar population, is necessary for the interpretation of the integrated light of high redshift galaxies.

The relevance of Galactic nuclear bulge giants for stellar synthesis models of early type galaxies was demonstrated by Whitford (1978); he showed that the integrated optical light of the bulge closely resembles that of early-type galaxies and of the bulges of other spirals. When bulge stars are included in models for galaxies the predicted energy distribution comes much closer to that of a typical real galaxy than any previous models (Frogel & Whitford 1987, Frogel *et al.* 1990, Terndrup *et al.* 1990, hereafter referred to as Papers I, II, and III, respectively). This remains true in spite of more recent work that finds that the mean metallicity of stars in Baade's Window is somewhat below solar and, therefore, below the mean metallicity of luminous E galaxies (McWilliam & Rich 1994; Houdashelt 1996; Tiede, Frogel, & Terndrup 1996)

Impey *et al.* (1986) presented the first high sensitivity survey of elliptical galaxies at 10 μm. They found significant – about a factor of two – excess emission (i.e., over and above that expected from normal stellar photospheres) at 10 μm in one-third of the galaxies in their sample and convincingly argued that this excess is actually present in most galaxies in their sample. Soifer *et al.* (1986) analyzed IRAS observations of the nuclear bulge of M31 over a region of about the same relative size as that interior to Baade's Window (at $b = -4°$ in the Galaxy) and found significant excess emission at 12 μm. Both Impey and Soifer *et al.* attributed this excess emission to thermal radiation from circumstellar dust associated with cool giants. In Paper I Whitford and I constructed a simple stellar synthesis model based on our observations of Baade's Window M giants, including observations we had made at 10 μm. *The predicted colors of this model at K, L and 10 μm were statistically indistinguishable from the observed colors of most of the E galaxies in Impey et al.'s (1986) sample and from the colors of the bulge of M31* (Soifer *et al.*1986).

While consistent with the general conclusion of Impey *et al.* (1986) and Soifer *et al.* (1986), namely that the origin of the 10 μm *excess* emission in the E galaxy sample and in M31 was circumstellar dust associated with late type stars, the simple model constructed in Paper I revealed that only a small fraction of the *total* emission at 10 μm inferred to come from bulge stars could be attributed to the *latest* M giants or to IRAS sources. Instead, the bulk of the 10 μm emission was shown to originate in M stars of intermediate spectral type (Table 5 of Paper I); nearly 90% of it coming from stars of type M7 and earlier. Since all M giants account for 65% of the total 10 μm emission in the model of Paper I (the remaining 35% comes from all other, ordinary stars), it was straight forward to demonstrate that if just the *excess* (as defined above) 10 μm emission is considered, over 80% of this is due to the M5-7 stars that individually have relatively weak excess emission. The contribution to this excess emission from the few IRAS sources identified in Baade's Window is ~10%.

The consistency arguments summarized above lead one to the conclusion that the stellar population of the bulge of the Milky Way, and by inference that of the bulges of other spiral galaxies and of elliptical galaxies, does not contain a *detectable* population of dust-enshrouded stars that emit primarily in the thermal IR (i.e., longward of 5 μm). In the case of the Milky

Way, images from the Diffuse Infrared Background Experiment (DIRBE) onboard the *Cosmic Background Explorer* (COBE) satellite offer striking qualitative evidence for this conclusion (Hauser 1993; Arendt *et al.* 1994; Weiland *et al.* 1994). The 12 and 25 µm images are dominated by the Zodiacal light and emission from the inner disk of the Galaxy; the latter is probably dominated by dust in the various components of the interstellar medium (ISM), although some of it will come from unresolved stars. No bulge is visible at these wavelengths. The images taken at wavelengths from 1.25 to 4.9 µm, on the other hand, are dominated by unresolved starlight; the bulge of the Galaxy is clearly visible. The colors of the bulge component are consistent with reddened starlight. Dwek *et al.* (1995) have used these shorter wavelength DIRBE data to characterize the structure of the bulge and to estimate its total luminosity and mass.

In spite of the strong dominance of light from the Galactic disk at 12 µm, it is still important to attempt a determination of the total emission from the Galactic bulge at this wavelength for comparison with observations of individual stars and of the integrated light from other galaxies. In this paper I describe such an estimate based on observations made by the InfraRed Astronomical Satellite (IRAS). Given the fact that IRAS' spatial resolution was nearly two orders of magnitude higher than that of DIRBE one might think that the 12 µm emission from the bulge could be measured by adding up the point sources listed in the IRAS Point Source Catalogue, the PSC. Such an approach, though, would suffer from two fundamental flaws – crowding and lack of detection of faint sources. Even at b = –4°, confusion problems are evident for the 12 µm sources in the PSC (Paper I). Furthermore, the IRAS PSC is a magnitude limited sample and thus contains only the very brightest part of the luminosity function for cool stars (Papers I and III). Taken together, these problems mean that an increasingly larger percentage of objects with excess emission will not be included in the PSC as the latitude decreases; hence, estimates of the total 12 µm surface brightness of the Galactic bulge based on the objects in the PSC will be systematically low by larger and larger amounts. Some efforts made to compensate for unresolved fainter sources have been discussed by Rowan-Robinson & Chester (1987) and Habing (1988). Finally, as mentioned above, at low latitudes emission at 12 µm is dominated by warm dust in the Galactic plane.

To get around problems caused by source confusion and the flux limited nature of the PSC, the analysis in this paper is based on an IRAS data product that was known as the Super Skyflux (SSF) data set. This data product was constructed by scientists at the Infrared Processing and Analysis Center (IPAC) in Pasadena[1] (IPAC). These images gave the *total* flux measured by IRAS in each pixel at each of the 4 IRAS wavelengths, corrected for the contribution from Zodiacal light. Hence, the images should include all emission from all point sources regardless of their individual luminosities in addition to any extended emission. The remainder of this paper presents an analysis of the SSF images that gave the 12 µm surface brightness as a function of latitude in the Galactic bulge for fields within which extensive ground based surveys and observations of individual stars have been carried out (Papers I, II, and III). The surface brightness determinations are then compared with model predictions based directly on the stellar observations. As will be shown below, the major problem that had to be overcome in this procedure was to reliably estimate the contribution of Galactic disk emission to SSF 12 μm images in the direction of the bulge. As noted above, the disk emission most likely arises primarily from the dust component of the ISM as well as from unresolved stars (including those with circumstellar shells). Unless stated otherwise, reference to "disk emission" in the remainder of this paper will refer to the sum of both such sources of emission that are strongly concentrated to the Galactic plane.

---

[1]The analysis to be described in this paper was carried out in late 1990; thus it is based on the IRAS data products and nomenclature as they existed at that time. The SSF data product has been superceded by the IRAS Sky Survey Atlas (ISSA). Improvements which may have been effected between the two data products, particularly with regards to the level of removal of zodiacal light and of "striping", would not alter the conclusions of this paper in any substantial way.

## 2. ANALYSIS OF THE IRAS SUPER SKYFLUX MAPS

The SSF was a representation of the entire sky as seen by IRAS smoothed to a resolution of 4 arcmin. The data product consisted of images 8° on a side with an overlap between neighboring images of 1 to 2°. Compared to previous large scale IRAS maps, SSF employed an improved model for the Zodiacal light to facilitate its removal, and an improved absolute flux calibration. In addition, a major effort was expended to minimize the effects of the "stripes" that can be seen in some two-dimensional representations of IRAS data (N. Gautier, private communication 1990). The techniques used minimized but did not entirely eliminate striping perpendicular to the ecliptic due to residuals from the removal of Zodiacal light. A somewhat earlier attempt to accomplish the same goals was a product called "BIGMAP" (W. Rice, private communication, 1990). In the course of my analysis I was able to show that the 12 and 25 μm surface brightness maps from BIGMAP and SSF near the galactic center, arguably the most difficult region to be dealt with by either data product, agreed to within a few percent.

SSF images at 12 μm were used to form a mosaic that covered a region extending ±34° in Galactic longitude and ±18° in Galactic latitude centered on the Galactic center. Programs available from IPAC made it possible to transform the images from ecliptic to Galactic coordinates. This mosaic closely resembled the 12 μm DIRBE image shown in Fig. 1 of Arendt *et al.* (1994) or the composite image contained in Fig. 2 of Hauser (1993). Inspection of these figures immediately points up the main problems facing this analysis: First, aside from the Zodiacal light, the only source of emission easily visible at 12 μm is the Galactic disk. The contrast between the 12 μm image and those at 1.2 to 3.5 μm (see either the Hauser or the Arendt et al. articles) is striking. The shorter wavelength images are dominated by starlight and show a well defined, flattened bulge with a thin disk extending on either side. On the other hand, no bulge is visible in the longer wavelength images. The second problem is the Zodiacal light. For a solar elongation angle of 90°, its intensity at 12 μm measured on the image in Arendt *et al.* is between 60 and 100 MJy/sr in the direction of the parts of the Galactic bulge considered in this paper. As will be shown below (Figures 1 and 2), the 12 μm intensity observed by IRAS in the direction of the bulge (after subtraction of Zodiacal light) is between 5 and 50 times less than this while the estimated component due to the bulge alone is down another factor of 5! The strength of the Zodiacal light is a strong function of solar elongation angle at the time of the observation; it can vary by as much as a factor of 3 for a change in elongation angle of 10°. Since the IRAS data taking procedure consisted of a series of sequential scans of strips of the sky each with a potentially different elongation angle, even small uncertainties in the removal of the Zodiacal light component can cause the striping effect in the final data that was mentioned earlier.

The problem, then, is how, on the 12 μm SSF map, to separate out whatever weak bulge emission may be present (independent of its source) from the overwhelming disk emission. The approach I took to resolving this problem was designed to result in a reasonable estimate for an upper limit to the 12 μm bulge emission.

Along any line of sight to the Galactic bulge there will be a contribution to the observed surface brightness from disk emission. A reasonable lower limit to the disk emission at a specific latitude within the bulge may be obtained by interpolating between its strength as observed at the same latitude but on either side of the bulge. This estimate could then be refined with a model that predicts how much of an increase is expected as one goes from outside to inside the bulge at constant latitude. Note that for the latitudes under consideration, the increase in extinction at 12 μm as one approaches $l = 0°$ is negligible.

It is first necessary to decide on a size for the bulge. I will consider 4 different estimates. Blanco & Terndrup (1989) mapped the distribution of M5 giants towards the Galactic bulge. They found that at a b = −6° the surface density of these giants is approximately flat for $\ell$ extending to ±5° on either side of the minor axis. In addition for $|\ell|$ between 15° and 25° (their survey limit) from the minor axis, they found that the contribution of the bulge to the star counts is negligible. Based on source counts from the IRAS PSC Habing *et al.* (1985) found no evidence for a bulge component at $|b|$ >10°, probably no evidence at $\ell$ outside of ±10° on either side of the bulge, and certainly no evidence for $\ell$ outside of ±15°. Van der Veen and Habing (1990) refined Habing *et al.*'s criteria for separating bulge and disk sources, but ended up with essentially the same values for the size of the bulge. Furthermore, their data, in agreement with the result of Blanco & Terndrup, showed that the number density of objects in the bulge is pretty flat at constant latitude at distances as great as ±6 ° in longitude from the minor axis. Finally, the COBE images (*e.g.* Hauser 1993; Dwek *et al.* 1995) show a well defined bulge in the *K* band with a linear extent of ±10° in latitude and a few degrees more than this in longitude. Dwek *et al.*'s contour maps confirm that because of the flattened shape of the bulge, the surface density is indeed fairly uniform for $|b|$ ≤ 5° and for $|\ell|$ ≤ 6°.

Based on these results, I will assume that the bulge has a uniform stellar surface density for $|b|$ ≤ 5° and $|\ell|$ < 6°, and that for $\ell$ in the range 15° < $|\ell|$ < 30° we are looking at a pure disk population so that this latter region can be used to estimate the disk contribution to the bulge emission. Finally, note that Blanco & Terndrup's (1989) data show that the contribution of the disk to star counts in the region ±6° on either side of the minor axis should be about 1.4 × what it is in the 15° < $|\ell|$ < 30° region.

Grism surveys for M giants, and subsequent near-IR photometry of them, were carried out in regions ~24′ in diameter along the minor axis of the bulge at latitudes of −3°, −4°, −6°, −8°, −10°, and −12° (Papers I and II). As discussed in Blanco, McCarthy, & Blanco (1984) and in Papers I and II the grism surveys are complete, i.e., a negligible fraction of M giants was missed since the faintest such stars are about two magnitudes brighter than the survey limit; the near-IR photometry is of an unbiased selection of the M giants in each of these fields. Later in this paper I will predict the 12 μm flux based on the integrated properties of the M giants in these 24′ diameter fields. Therefore, it was convenient to rebin the SSF mosaics to a pixel size of 12 arcmin square.

Cuts through the rebinned 12 μm SSF mosaics at the latitudes of the grism surveys, and at −2°, are shown in the bottom panel of Figure 1. Similar cuts at positive latitudes are shown in the bottom panel of Figure 2. For analysis purposes, each cut was divided into three segments, one that extended 6° on either side of the minor axis, corresponding to the region of approximately constant surface density in the bulge, and ones that extended from 15° to 30° from the minor axis on both sides. These latter two cuts will be used to derive the best estimate for the disk surface brightness in the bulge for each cut. In order to minimize the effects of bright individual point sources, many of which are probably foreground stars, further smoothing of the cuts in the bottom panels of Figs. 1 and 2 was done by replacing all pixels in the observed cuts that differed by more than ±0.11 MJy/sr from neighboring pixels by an average of the nearest neighbors. This procedure was iterated twice. As pointed out elsewhere in this paper, observations of individual IRAS point sources in bulge fields indicates that they contribute only a few percent to the total flux in any given field. These resulting smoothed cuts are displayed in the upper parts of the figures. The broad hills and valleys that can be followed over several adjacent cuts in the figures result from the striping effects referred to earlier. Thus, the smoothing process used should not significantly affect the flux values measured from the cuts.

Columns 2 and 4 of Table 1 are the median values measured directly on the smoothed 12 µm cuts at the latitudes in col. 1 and between 15 and 30 degrees on either side of the minor axis. Col. 3 lists the median values between ±6° on either side of the minor axis, again measured directly from the smoothed cuts. Col. 5 lists the difference between col. 3 and the average of cols. 2 and 4, i.e., these values should be the surface brightness values of the bulge at 12 µm and at the specified latitudes with the disk contribution removed. Since this procedure assumes that the surface brightness of the disk remains constant at a given latitude, the disk corrected values for the bulge surface brightness in col. 5 are probably upper limits to the true values. Col. 6 gives the surface brightness values in col. 5 in more convenient units. The fact that some of the estimates for pure bulge surface brightness at 12 µm are negative arises from the striping, i.e., from the inability to exactly account for the extended zodiacal light emission. Of course, the imperfect removal of zodiacal light can also cause some of the surface brightness values to be "too positive." The values in cols. 5 and 6 would not be significantly different if averages instead of medians were used in calculating the disk and bulge contributions.

### 3. MODEL PREDICTIONS

Table 2 presents the comparison of disk-subtracted 12 µm bulge surface brightness values with predicted values based on the observed numbers and photometric properties of M stars in fields on or very close to the Galaxy's minor axis (Papers I & II). The observations of the M giants of course include any excess emission from these stars that could arise from circumstellar dust. Columns 2 & 3 of Table 2 repeats the values from col. 6 of Table 1 while col. 4 is the average of these two columns. The 1 σ standard deviations are based on the scatter in the 4 disk values and the two bulge values needed to calculate the averages in this column. In spite of the fact that the predicted values are in much better agreement with the values observed at negative latitudes, we will discuss only the averages since as indicated above, there is no *a priori* reason to assume that the values at negative latitudes are any better after correction for zodiacal light than the values at positive latitudes.

Column 5 of Table 2 lists the predicted 12 µm surface brightness values based on Papers I and II. Figure 3 compares these predicted values (solid line) with the average values in col. 4. The predicted values were calculated as follows. Tables 4 and 5 of Paper I give the integrated colors and magnitudes of the M giants in Baade's Window, the $b = -4°$ field, and the relative contribution of the different types of M giants to the integrated light from all of these stars. Tables 7 and 8 of Paper II give similar information for all of the other observed fields between $b = -3°$ and $-12°$.

Although integrated $K - [10]$ values ([10] stands for the magnitude at 10 µm) have been determined only for the Baade's Window field, this color can be estimated quite accurately for the remaining fields in the following manner: Table 5 of Paper I shows that each M subtype makes the same fractional contribution to the flux in the *J* and to the *L* bands but that this fractional contribution changes substantially with M subtype. Furthermore, the mean *JHK* colors of each subtype do not change much with latitude. Thus I assume that any observed changes in integrated color with latitude will be due primarily to the changes in the *mix* of subtypes. Based on the observations of stars in Paper II with data at *L* and at 10 µm, the *J–K*, *K*–[10] and *J–K*, *K–L* color-color relations appear to be the same for all stars observed, whether or not they are in Baade's Window. Thus, I further assume that changes in the integrated *K–L* and *K*–[10] colors will scale with the integrated *J–K* color. For *b* between –3° and –8° $(J-K)_0$ seems to be constant, changing only between 1.05 and 1.08. Therefore, for these fields I use an integrated *K*–[10] = 0.8, as determined for Baade's Window (Table 4, Paper I). For the –10 and –12° fields $(J-K)_0 \approx 0.98$. In Paper I it is argued that M giants contribute ~65% of the 10 µm light with the remaining 35% coming from all other ordinary, non-M stars. The spectral mix and reduced metallicity of the outer two fields (Tiede, Frogel, & Terndrup 1996) suggests that the contribution of M giants to the integrated light of these fields will be ~10% less than for the other fields. Thus, I estimate that for the –10 and –12° fields the *K*–[10] integrated color is 0.7. The flux from a [10]=0 star is taken to be 37 Jy. Each of the fields for which integrated magnitudes are given in Papers I and II have diameters of 24.4′. The last assumption I make is that the predicted flux at 10 µm should be about equal to that at 12 µm[2]. These values are given in col. 5 of Table 2. Note that because of the small numbers of M giants in the 10 and 12° fields, the model predictions will be subject to considerable uncertainty; indeed, the surface brightness is predicted to be greater at the higher latitude field, an obviously non-physical result.

Finally, Table 2 also contains predictions for a field at ***b*** = –2 °. This prediction is based on an extrapolation of the observed dependence of *K* surface brightness with latitude determined from the observations of the fields with *b* between –3 and –12°.

## 4. CONCLUSIONS

For each of the bulge fields listed in Table 2, the predicted 12 µm surface brightness is somewhat greater than the average of the observed surface brightness at the corresponding positive and negative galactic latitudes (see Fig. 3). Since the correction I applied to the IRAS bulge emission for the disk contribution is probably a lower limit, the true difference could be greater. On the other hand, since I assumed the same flux between 10 and 12 µm whereas a plausible alternative would be to have assumed nearly zero color, the predicted values for the 12 µm emission could be reduced by about 30%. The uncertainties in the calculated 12 µm IRAS surface brightness values are large because the correction for disk emission is of comparable size to the total emission observed at each latitude, and because even small uncertainties in the correction for Zodiacal light will be large compared to any of the values of interest. Nevertheless, one of the main conclusions of this paper is the following: On the basis of IRAS observations there is no evidence for significant 12 µm emission in the galactic bulge other than that expected from normal stars, especially M giants; nearly all of the latter have been identified optically. This normal emission includes the small excess emission observed from these M giants (Paper I). More specifically, this result rules out the possibility of an unresolved, *significant* population of dust enshrouded stars in the bulge with extensive circumstellar envelopes. It also rules out the possibility of any *extended* source of 12 μm emission in the bulge as could arise from a dusty ISM of density greater than which would be expected from the intermixing of the disk and bulge in this region.

---

[2] This assumption takes into account the small excess of emission that is observed. If the [10–12] µm *color* were close to 0, as would be the case for black body emission, then the model predicted fluxes would have to be reduced by a factor of 1.3 which, as will be seen, would bring these values into better agreement with the IRAS observations.

Frogel & Whitford (1987) identified 7 IRAS point sources in a region centered on Baade's Window (*b* = – 4°), but slightly larger in diameter. Taking their fluxes from the IRAS PSC, I calculate that these 7 stars would have a surface brightness of 46 Jy/□°, only one-fifth of the value derived from the 12 μm images and one-tenth of the model predicted value.

The results of this paper combined with those of Paper I imply that the strength of the 12 μm emission from the Galactic bulge as deduced from IRAS observations is consistent with the observed small excess of emission in elliptical galaxies and the bulges of other spirals. Furthermore, these results lead to the conclusion that if the cool giants in the galactic bulge are typical of those to be found in elliptical galaxies, then 80% of the factor of two excess emission observed in ellipticals (Impey, Wynn-Williams & Becklin 1986) is due to circumstellar dust associated with M5-7 giants while the remaining 20% is from M8-9 giants. No exotic stellar types are required to explain the excess emission.

My research at IPAC with IRAS data was supported by NASA grant NAG 5-1367, in the Astrophysics Data Program. I thank the many people at IPAC who assisted me with this program. I also thank my OSU colleagues Kris Sellgren and Don Terndrup for suggestions which improved this paper.

# REFERENCES

Aaronson, M. 1977, Ph.D. thesis, Harvard Univ.
Arendt, R. G. et al. 1994, ApJL, 425, L85
Blanco, V. M., McCarthy, M. F., & Blanco, B. M. 1984, AJ, 89, 636
Blanco, V. M., & Terndrup, D. M. 1989, AJ, 98, 843
Frogel, J. A., & Whitford, A. E. 1987, ApJ, 320, 199 (Paper I)
Frogel, J. A. 1988, ARA&A, 26, 51
Frogel, J. A., Persson, S. E., Aaronson, M., & Matthews, K. 1978, ApJ, 220, 75
Frogel, J. A., Terndrup, D., Blanco, V. M., & Whitford, A. E. 1990, ApJ, 353, 494 (Paper II)
Habing, H. J. 1988, A&A, 200,
Habing, H. J., Olnon, F. M., Chester, T., Gillett, F., Rowan-Robinson, M., & Neugebauer, G. 1985, A&A 152, L1
Hauser, M. G. 1993, in Back to the Galaxy, ed. S. S. Holt & F. Verter (New York: AIP), 201
Houdashelt, M. L. 1996, Ph.D. thesis, Ohio State University
Impey, C. D., Wynn-Williams, G., & Becklin, E. E. 1986, ApJ, 309, 572
McWilliam, A., & Rich, R. M. 1994, ApJS, 91, 749
Rowan-Robinson, M., & Chester, T. 1987, ApJ, 313, 413
Soifer, B. T., Rice, W. L., Mould, J. R., Gillett, F. C., Rowan-Robinson, M., & Habing, H. J. 1986, ApJ, 304, 651
Terndrup, D. M., Frogel, J. A., & Whitford, A. E. 1990, ApJ, 357, 453 (Paper III)
Tiede, G. P., Frogel, J. A., & Terndrup, D. M. 1995, AJ, 110, 2788
Tinsley, B. M. 1972, ApJ, 178, 319.
van der Veen, W. E. C. J., & Habing, H. J. 1990, A&A, 231, 404
Weiland, J. L. et al. 1994 ApJL, 425, L81
Whitford, A. E. 1978, ApJ, 226, 777

# FIGURE CAPTIONS

Fig. 1 — (bottom) The unsmoothed cuts at negative latitudes from the SSF plates rebinned to a resolution of 12′ as described in the text. (top) The smoothed cuts at negative latitudes.

Fig. 2 — (bottom) Similar to Fig. 2 except these cuts are at positive latitudes. (top) The smoothed cuts at positive latitudes.

Fig. 3 — The points are the averages of the observed 12 µm surface brightness values at positive and negative latitudes from Table 2. The solid line shows the predicted values, also from Table 2.

## TABLE 1

**MEDIAN 12 μm VALUES FROM SUPER SKY FLUX MOSAICS**

| b (°) (1) | disk: $\ell$ = +30° to +15° (MJy/sr) (2) | bulge: $\ell$ = +6° to –6° (MJy/sr) (3) | disk: $\ell$ = –30° to –15° (MJy/sr) (4) | bulge – disk (MJy/sr) (5) | bulge – disk (Jy/□°) (6) |
|---|---|---|---|---|---|
| -12 | 0.43 | 1.13 | 0.71 | 0.56 | 170 |
| -10 | 1.09 | 1.16 | 0.85 | 0.20 | 60 |
| -8 | 1.37 | 1.23 | 1.03 | 0.04 | 11 |
| -6 | 1.69 | 2.32 | 1.68 | 0.64 | 196 |
| -4 | 3.00 | 4.05 | 3.43 | 0.84 | 255 |
| -3 | 4.26 | 6.14 | 5.48 | 1.27 | 388 |
| -2 | 7.29 | 10.84 | 9.38 | 2.50 | 763 |
| 2 | 8.36 | 10.52 | 9.15 | 1.76 | 538 |
| 3 | 5.09 | 6.14 | 5.68 | 0.76 | 232 |
| 4 | 3.97 | 4.81 | 4.15 | 0.75 | 228 |
| 6 | 2.65 | 2.19 | 2.61 | -0.43 | -132 |
| 8 | 2.48 | 1.37 | 1.87 | -0.80 | -245 |
| 10 | 2.12 | 0.91 | 1.47 | -0.88 | -269 |
| 12 | 2.17 | 0.69 | 1.11 | -0.96 | -291 |

# TABLE 2

**COMPARISON OF 12 μm SUPER SKY FLUX VALUES WITH PREDICTIONS**

| |b| | bulge − disk | | | model prediction |
|---|---|---|---|---|
| | $b < 0°$ | $b > 0°$ | (average) | |
| (°) | (Jy/□°) | (Jy/□°) | (Jy/□°) | (Jy/□°) |
| (1) | (2) | (3) | (4) | (5) |
| 12 | 170 ±44 | -291 ±162 | -61 ±135 | 21 |
| 10 | 60 ±36 | -269 ±99 | -104 ±93 | 12 |
| 8 | 11 ±53 | -245 ±93 | -117 ±98 | 65 |
| 6 | 196 ±2 | -132 ±6 | 32 ±86 | 112 |
| 4 | 255 ±66 | 228 ±27 | 241 ±140 | 370 |
| 3 | 388 ±187 | 232 ±91 | 310 ±96 | 446 |
| 2 | 763 ±318 | 538 ±120 | 650 ±152 | 1100 |

13